\documentstyle[12pt,aaspp4]{article}

\newcommand{\uc}{\rm U Com}

\newcommand{\lsun}{log $L/L_{\odot}\,$}
\newcommand{\msun}{$M/M_{\odot}\,$}

\begin{document}

\title{Theoretical models for Bump Cepheids}

\author{Giuseppe Bono\altaffilmark{1}, Vittorio Castellani\altaffilmark{2}, 
Marcella Marconi\altaffilmark{3}}

\affil{1. Osservatorio Astronomico di Roma, Via Frascati 33,
00040 Monte Porzio Catone, Italy; bono@mporzio.astro.it}
\affil{2. Dipartimento di Fisica Universit\`a di Pisa, Piazza Torricelli 2,
56100 Pisa, Italy; vittorio@astr18pi.difi.unipi.it}
\affil{3. Osservatorio Astronomico di Capodimonte, Via Moiariello 16,
80131 Napoli, Italy; marcella@na.astro.it}

\begin{abstract}
We present the results of a  theoretical investigation 
aimed at testing whether full amplitude, nonlinear, convective models 
account for the I-band light curves of  Bump Cepheids in the Large 
Magellanic Cloud (LMC). We selected two objects from the OGLE sample 
that show a well-defined bump along the decreasing (short-period) and 
the rising (long-period) branch respectively. We find that current 
models do reproduce the luminosity variation over the entire pulsation 
cycle if the adopted stellar mass is roughly 15\% smaller than predicted 
by evolutionary models that neglect both mass loss and convective core 
overshooting.  
Moreover, we find that the fit to the light curve of the long-period 
Cepheid located close to the cool edge of the instability strip 
requires an increase in the mixing length from 1.5 to 1.8 Hp. 
This suggests an increase in the efficiency of the convective transport 
when moving toward cooler effective temperatures. Current pulsation 
calculations supply a LMC distance modulus ranging from 18.48 to 18.58 mag.
\end{abstract}

\keywords{Cepheids -- Magellanic Clouds -- stars: distances -- 
stars: evolution -- stars: oscillations}  

\pagebreak 
\section{Introduction} 

Classical Cepheids are key objects for estimating stellar distances and for 
calibrating secondary distance indicators necessary to 
evaluate the Hubble constant (Ferrarese et al. 2000; Saha et al. 1999). 
From a theoretical point of view Cepheids 
can also be regarded as fundamental physical laboratories to assess both 
the accuracy of the input physics (Simon 1989) and the reliability of the 
physical assumptions adopted in modeling radial pulsations 
(Bono, Marconi, \& Stellingwerf 1999, hereinafter BMS). 
Thus, any new test concerning the accuracy of theoretical 
predictions appears of paramount relevance.  Moreover, this effort is mandatory to improve our 
confidence on Cepheids as a robust rung of the cosmic distance 
ladder and  to disentangle relevant open 
questions such as the dependence of the  Cepheid distance scale 
either on metal content (Bono et al. 1999; Bono, Castellani 
\& Marconi 2000a; Caputo et al. 2000; Laney 2000; Storm et al. 2000) 
or on blending (Mochejska et al. 2000).

In this context, one has to remind that the current theoretical framework 
based on nonlinear, convective pulsation models, is facing with a  
fundamental problem, i.e. the calibration of the turbulent convection 
(TC) model adopted to account for the coupling between pulsation and 
convection (Castor 1968; Stellingwerf 1982). The approach originally 
suggested by Bono \& Stellingwerf (1993, 1994, hereinafter BS) relies 
on the combination of leading physical arguments and empirical constraints 
on the topology of the RR Lyrae instability strip in Galactic globulars. 
Such a theoretical scenario accounts for some relevant properties of 
radial variables in the Cepheid instability strip (Bono et al. 1997a,b; 
BMS; Bono, Caputo, \& Marconi 2001). However, the predicted RR Lyrae 
light curves close to the low-temperature instability edge are somehow 
at variance with empirical data (Kovacs \& Kanbur 1998).
This appears as a disturbing evidence, since empirical light curves 
supply tighter constraints on the accuracy of theoretical predictions 
than mean magnitudes and/or pulsational periods. 
 
In a recent paper (Bono, Castellani,\& Marconi 2000b, hereinafter BCM)
we have approached such a problem and successfully reproduced the 
luminosity variation over a full pulsation cycle of the field, first 
overtone RR Lyrae \uc. At the same time, we also constrained 
the adopted TC  model, and indeed in that paper we found that the Bump 
located just before the luminosity maximum can be reproduced by nonlinear 
models only by assuming a vanishing overshooting efficiency at the 
boundaries of the convective unstable regions. 
To explore the general reliability of this finding, we undertook a further 
investigation aimed at checking whether this TC model can also account 
for Cepheid light curves. To this purpose we selected two Bump Cepheids 
from the OGLE database (Udalski et al. 1999) of the LMC, with the Bump 
either along the decreasing (OGLE194103, short-period) or along the 
rising (OGLE56087, long-period) branch of the light curve. 
Table 1 lists the empirical data of the selected objects.

The reason 
for this choice follows from the evidence that both the amplitude and 
the phase of the Bump are crucial features to nail down the predicting 
power of theoretical models. Moreover, both stars present quite similar 
apparent magnitudes, and therefore the period difference suggests that 
the short-period Cepheid is located close to the blue (hot) edge, while 
the long-period one close to the red (cool) edge of the instability strip. 
When moving from hotter to cooler effective temperatures the thickness 
of the convective unstable region, and in turn the efficiency 
of the convective transport increases. This means that the envelopes of 
the two Bump Cepheids present substantially different physical structures.  
Finally, the intimate nature of the Hertzsprung progression 
is still controversial (Bono, Marconi, \& Stellingwerf 2000c), 
and a reliable fit to empirical light curves can supply new insights into 
the physical mechanisms that govern the occurrence of this phenomenon.

\section{Theoretical best fit models}

To fit the empirical light curve of \uc, BCM computed several 
iso-period sequences of nonlinear models at fixed chemical composition.  
For each given assumption on the stellar mass, the iso-period sequence 
was constructed by changing a single input parameter, namely the 
effective temperature, since the luminosity  was constrained by the 
observed period.  
However, the fit to Bump Cepheid light curves will be performed by 
adopting a different and stronger constraint. At first we 
undertake the fit of one of the two variables according to suitable 
Mass-Luminosity (ML) relations based either on canonical or on mild 
overshooting evolutionary models (BMS; Girardi et al. 2000), both 
neglecting mass-loss. This approach obviously decreases the 
degrees of freedom of the problem. In fact, for each selected luminosity 
we adopt the stellar mass as given by the  ML relation, thus only 
one effective temperature accounts for the constraint on the observed 
period. If and when the best fit of the light curve of the 
former Cepheid is accomplished, we derive the stellar parameters 
(L, M, Te), and in turn an estimate of the LMC distance modulus. 
According to this result, the fit to the light curve of the latter 
object will be accomplished under the even stronger constraint to  
adopt the same ML relation and the same LMC distance modulus 
already derived for the former object.  

For both stars we adopted a chemical composition typical of LMC 
Cepheids (Y=0.25, Z=0.008). We considered at first the short-period 
Cepheid OGLE194103 and constructed various iso-period ($P=8.7 d$) 
sequences at different luminosity levels by adopting the canonical 
ML relation used in previous papers (BMS; Bono et al. 1999; 
Bono et al. 2000d). A glance at 
these iso-period sequences disclosed that the shape of the light curve 
is mainly governed by the stellar parameters L, M, and Te, whereas the 
luminosity amplitude sensitively depends on the adopted mixing length 
parameter. However, we found that under the quoted assumptions theoretical 
models appear unable to reasonably fit observational constraints. Several 
numerical experiments were performed to test the dependence of both 
pulsation amplitude and shape of the light curve on the free parameters 
adopted in the TC model (BS), but the shape of the predicted light curve 
was still at variance with the observed one.  

Therefore we decided to test models with a noncanonical ML relation. 
To mimic the luminosity predicted by evolutionary models that account 
for mild convective core overshooting (Girardi et al. 2000), we increased 
the luminosity of canonical models, for each stellar mass, by 0.25 dex 
(Chiosi et al. 1993).  We repeated once again the quoted experiments and 
now we found that our basic TC model gives a reasonable fit to the 
OGLE194103 light curve for the couple of parameters  \lsun=3.55 
(\msun=5.57), and $T_e=5660$ K, and our canonical mixing length value 
i.e. {\em l}=1.5 Hp (BS94). Note that the value of the effective temperature 
refers to the static linear model. The temperature variation along the 
pulsation cycle of the best fit nonlinear model is equal to 850 K. 
The agreement between theory and observations  appears quite satisfactory 
over the entire pulsation cycle (see Fig. 1) and accounts for the phase 
and the amplitude of the Bump. 

The same approach has been applied to fit the light curve of the 
long-period Bump Cepheid (OGLE56087, P=13.6 d). In particular, we 
adopted the same ML relation and the same distance modulus (DM=18.48) 
we found for the short-period Cepheid.
The bottom panel of Fig. 1 does show that a remarkable agreement 
between theory and observations is obtained by adopting, \lsun=3.61 
(\msun=5.70), and $T_e=5160$ K, but now we were forced to assume 
{\em l}=1.8 Hp. 
In fact, the model constructed by adopting {\em l}=1.5 Hp (dotted line) 
presents the same secondary features but the luminosity amplitude is 
systematically larger, i.e.  $\Delta I$=0.88 against the observed 
0.51 mag. A glance at the data plotted in the bottom panel clearly 
shows the accuracy with which the best fit model does reproduce the 
well-defined dip of the observed light curve at $\phi\approx0.4$, 
soon after the Bump secondary maximum. Present computations 
thus suggest that the mixing length parameter {\em l} becomes larger 
when a star becomes cooler, and the convective unstable region becomes 
thicker, i.e. the efficiency of the convection increases more than 
predicted by the assumption of a constant mixing length.

The use of a slightly higher metal abundance Z=0.01, according to 
recent empirical evidence (Luck et al. 2000) that LMC Cepheids seem 
to cover a wide metallicity range ($0.006 \le Z \le 0.013$), does 
not improve the fit. The same outcome applies to plausible changes 
in the He content, and indeed a change from 0.25 to 0.26 does 
affect neither the shape of the light curve nor the luminosity 
amplitude.  

To investigate the robustness of current best fit solutions we also 
explored whether plausible changes in the input parameters still 
supply a reasonable fit to the observed light curves. The top panel
of Fig. 2 displays that a nonlinear model constructed by adopting 
\lsun=3.586 (\msun=5.7), $T_e$=5780 K, and a mixing length 
{\em l}=1.5 Hp properly fits the luminosity variation of the 
short-period Cepheid (OGLE194103). This finding implies a LMC 
distance modulus of 18.58 mag and the fit to the long-period Cepheid 
(OGLE56087) now requires: \lsun=3.65 (\msun=5.95), $T_e=5225$ K, and 
a mixing length {\em l}=1.85 Hp (see Fig. 2). No agreement was found 
outside this range of stellar luminosities.  

As a whole, previous numerical experiments support the evidence that 
the {\em pulsational} LMC distance modulus can be safely bracketed  
between 18.48 and 18.58 mag. Moreover, a decrease in the evolutionary 
mass between canonical and best fit models of the order of 15\% is, 
in any case, required. Finally, the variation of a single free parameter,
the mixing length, allows us to account for the shape of empirical Bump 
Cepheid light curves across the instability strip.

\section{Discussion}

According to the results given in the previous section, our pulsational
computations require a larger mixing length value when the star becomes 
cooler.  We note that this is not a very surprising result, since a mixing length 
roughly equal to 1.9 Hp is also required by the evolutionary models  
that best fit the red giant branch of Galactic globular clusters 
(Castellani 1999).
This notwithstanding, as already discussed in BCM, a substantial variation 
in the mixing length parameter causes a sizable change in the efficiency 
of the convective transport across the driving regions, and in turn in the 
temperature width of the instability strip.  
To estimate the dependence of the instability edges on {\em l} we 
constructed new models with the stellar mass of the long-period best 
fit model. We find that when moving from {\em l}=1.5 to 1.8 Hp the 
temperature width decreases from 1100 K to 800 K. This change is mainly 
due to a shift toward hotter effective temperatures of the red boundary 
and could be marginally at odds with empirical 
estimates. In fact, estimates by Pel \& Lub (1978, see their Fig. 3) 
do suggest that in this period range the strip is slightly larger than 
800 K. Unfortunately, we are not aware of any recent empirical estimates 
that can allow us to supply firm constraints on this observable. 
 
The pulsational scenario we are dealing with 
accounts for the empirical light curves of the selected Bump Cepheids 
provided that their stellar masses are 15\% smaller than predicted by 
canonical evolutionary models. This theoretical evidence, if taken at 
face value, could be due either to the occurrence of mass-loss among 
intermediate-mass stars, as originally suggested by Cox (1980), 
or to convective core overshooting, or both of them. Unfortunately, 
the pulsational approach adopted in this investigation does not 
allow us to discriminate between the two different hypothesis. 
Current results supply some support to the evidence recently brought 
out by Beaulieu et al. (2001) concerning the {\em discrepancy} between 
evolutionary and pulsational masses among Magellanic Cepheids. 
On the basis of a new approach  they found that evolutionary 
masses are up to $\Delta \log M/M_\odot = 0.1$ larger than pulsational 
ones. On the other hand, we found that for the two selected Bump Cepheids 
the discrepancy is of the order of 0.07 and 0.08 dex for the short and 
the long-period respectively.  

Note that Bono et al. (2001)
performed a detailed analysis of both pulsation and evolutionary
masses of Galactic Cepheids for which are available both accurate
empirical estimates of mean radii, distances, and individual reddenings.
Interestingly enough, they found that pulsational masses are
approximately 10\% smaller than evolutionary ones.
A similar result was found by Wood, Arnold, \& Sebo (1997) by fitting 
the light curve of a LMC Bump Cepheid (HV905). However, they found that 
the luminosity of this object was substantially higher than predicted 
by ML relations based either on canonical or on mild overshooting 
evolutionary models. 

To assess whether our best fit models support either the occurrence of a 
resonance between fundamental and second overtone or the Christy's echo 
mechanism (Bono et al. 2000c, and references therein) we finally constructed 
two new models by adopting the same input parameters of the models 
plotted in Fig. 1, but by perturbing the linear second overtone radial 
eigenfunctions. We find that both of them, after a transient phase 
lasting $\approx$200 (long-period) and $\approx$700 (short-period) 
cycles, undergo a mode switch from the second to the fundamental 
mode. The period ratios between second overtone and 
fundamental period are $P_2/P_0 (OGLE194103)$=0.523-0.528 and 
$P_2/P_0 (OGLE56087)$=0.495-0.503 respectively. For each object 
the former value was derived by adopting linear $P_2$ and $P_0$ periods,
while the latter one with linear $P_2$ and nonlinear $P_0$ periods.  
The values of these ratios still fall in the so-called {\em resonance 
region} (Simon \& Schmidt 1976). However, we  note that the physical 
structure of the best fit models does not show any nodal line 
(Bono et al. 1997c) along the pulsation cycle. This finding together 
with the good agreement between theoretical predictions and empirical 
data further strengthens the plausibility that the Bumps are triggered 
by the Christy's echo mechanism.

\acknowledgments
This work was supported by MURST-Cofin 2000, under the scientific project 
"Stellar Observables of Cosmological Relevance".

\pagebreak

\pagebreak
\begin{center}
\begin{tabular}{lcc}
\tablewidth{0pt}\\
\multicolumn{3}{c}{TABLE 1. Empirical data$^a$}\\
\hline
                 & OGLE194103       & OGLE56087  \\
\hline
R.A. (J2000)     &$05^h 06^m 33.96^s$           &$05^h 00^m 48.3^s$\\
DEC. (J2000)     &$-68^{\circ} 25^{\arcmin} 12.8^{\arcsec}$ &$-69^{\circ} 31^{\arcmin} 54.8^{\arcsec}$\\ 
Period (d)       &8.71306           &13.62635     \\
$m_V$  (mag)     &$14.68\pm0.01$   &$14.66\pm0.01$ \\
$m_I$  (mag)     &$13.90\pm0.01$   &$13.68\pm0.01$ \\
$\Delta_V$ (mag) &$0.80 $         &0.86           \\
$\Delta_I$ (mag) &0.46             &0.51           \\
\hline
\end{tabular}
\end{center}
\begin{minipage}{1.00\linewidth}
\noindent $^a$ Empirical data for the two selected Bump Cepheids 
according to Udalski et al. (1999).     
\end{minipage}

\pagebreak

\figcaption{Comparison between predicted (solid line) and observed (open 
circles) I-band light curves for the short-period (OGLE194103, top panel) 
and the long-period (OGLE56087, bottom panel) Bump Cepheid. The input
parameters of the best fit model are labeled.  By assuming a color
excess equal to E(B-V)=0.1 mag, the inferred true distance modulus
for the two objects is 18.48 mag. The dotted line plotted in the bottom
panel shows the light curve of a model constructed by adopting the same
input parameters of the best fit model but with a mixing length 
{\em l}=1.5 Hp. See text for more details.}

\figcaption{Same as Fig. 1, but the pulsation models were constructed
by adopting the labeled input parameters. By assuming a color excess 
equal to E(B-V)=0.1 mag, the inferred true distance modulus for the 
two objects is 18.58 mag.}

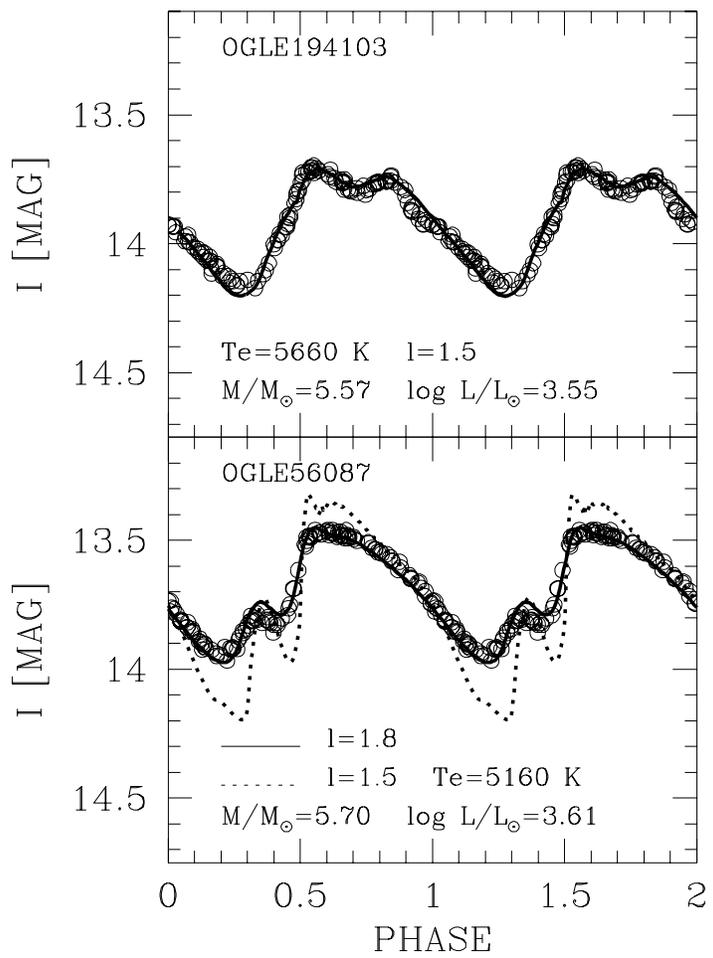
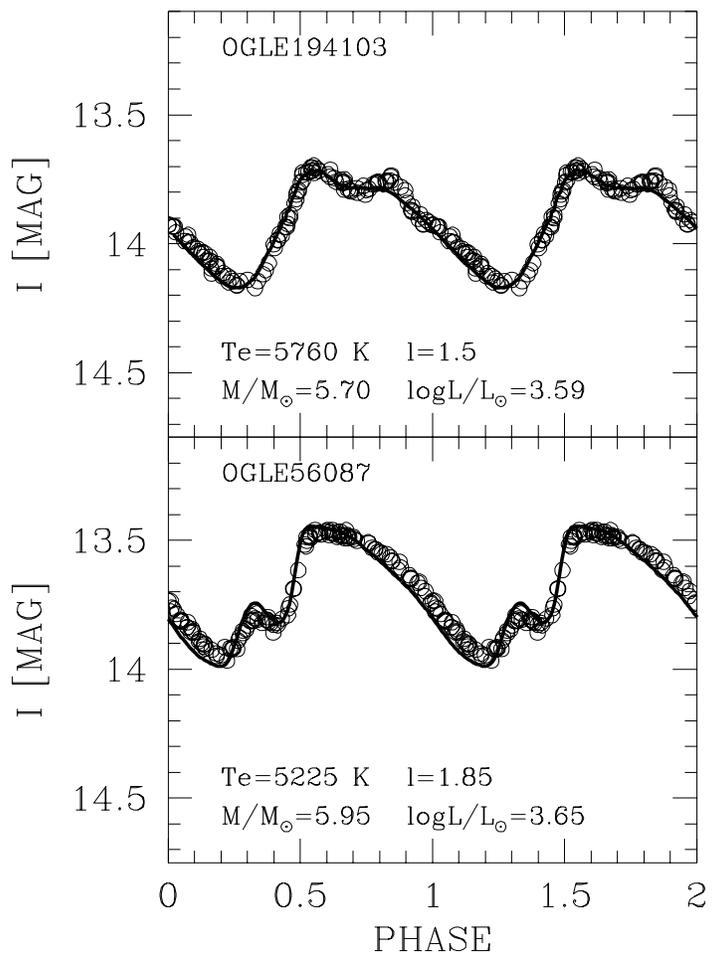
\end{document}